\documentclass[fleqn,usenatbib]{mnras}

\usepackage{newtxtext,newtxmath}

\usepackage[T1]{fontenc}
\usepackage{ae,aecompl}

\usepackage{graphicx}	
\usepackage{amsmath}	
\usepackage{amssymb}	
\usepackage[colorinlistoftodos]{todonotes}

\title[Long-term study of PSR\,B0950+08 with AARTFAAC]{Long-term study of extreme giant pulses from PSR\,B0950+08 with AARTFAAC}

\author[M. J. Kuiack et al.]{
Mark Kuiack,$^{1}$\thanks{E-mail: m.j.kuiack@uva.nl}
Ralph A.M.J. Wijers,$^{1}$
Antonia Rowlinson,$^{1,2}$\newauthor
Aleksandar Shulevski,$^{1}$
Folkert Huizinga,$^{1}$
Gijs Molenaar,$^{3}$
Peeyush Prasad,$^{1}$
\\
$^{1}$ Anton Pannekoek Institute for Astronomy, University of Amsterdam, Science Park 904, 1098 XH, Amsterdam, The Netherlands \\
$^{2}$ ASTRON, Netherlands Institute for Radio Astronomy, Oude Hoogeveensedijk 4, 7991 PD Dwingeloo, The Netherlands \\
$^{3}$ Department of Physics and Electronics, Rhodes University, PO Box 94, Grahamstown, 6140, South Africa \\
}

\date{Accepted XXX. Received YYY; in original form ZZZ}

\pubyear{2020}

\begin{document}
\label{firstpage}
\pagerange{\pageref{firstpage}--\pageref{lastpage}}
\maketitle

\begin{abstract}
We report on the detection of extreme giant pulses (GPs) from one of the oldest-known pulsars, the highly variable PSR\,B0950+08, with the Amsterdam-ASTRON Radio Transient Facility And Analysis Centre (AARTFAAC), a parallel transient detection instrument operating as a subsystem of the LOw Frequency ARray (LOFAR).
During processing of our Northern Hemisphere survey for low-frequency  radio transients, a sample of 275 pulses with fluences ranging from 42 to 177\,kJy\,ms were detected in one-second snapshot images.
The brightest pulses are an order of magnitude brighter than those previously reported at 42 and 74\,MHz, on par with the levels observed in a previous long-term study at 103\,MHz. Both their rate and fluence distribution differ between and within the various studies done to date. The GP rate is highly variable, from 0 to 30 per hour, with only two three-hour observations accounting for nearly half of the pulses detected in the 96\,h surveyed. It does not vary significantly within a few-hour observation, but can vary strongly one from day to the next.
The spectra appear strongly and variably structured, with emission sometimes confined to a single 195.3\,kHz subband, and the pulse spectra changing on a timescale of order 10\,min.  
\end{abstract}




\section{Introduction}

PSR\,B0950$+$08 was first reported as Cambridge Pulsed source 0950, CP0950 \citep{1968Natur.218..126P}, soon after the first pulsar, CP1919, was detected \citep{1968Natur.217..709H}.  CP0950 has a period of 0.253\,s and is very nearby, with a dispersion measure of only 2.97\,pc\,cm$^{-3}$ and a distance of 262$\pm$5\,pc \citep{Brisken+2002}, improved from an earlier measurement of 127$\pm$13\,pc by \cite{Gwinn+1986}. Its main pulse profile is single-peaked at higher frequencies but evolves to a classic double-peaked emission-cone shape at LOFAR frequencies \citep[e.g.,][and references therein]{2016A&A...586A..92P}, with a width of about 25\,ms at 60\,MHz. It also has a variable interpulse \citep{HankinsCordes1981}, which can in rare cases be as bright as the main pulse \citep{Cairns+2004}, but since our observations do not resolve the pulse at all, we cannot distinguish between the two and have to group their effects together. 

Already in its discovery paper CP0950 was reported to be highly variable in intensity, by more than a factor ten, much more so than other pulsars \citep[e.g.,][]{Cole+1970}. It was also soon recognised that pulsar pulse profiles could consist of sub-pulses \citep[e.g.,][]{Cole1970}. Specifically, \cite{Hankins1971} showed that CP0950 could have very fine and highly variable time structure, sometimes with very bright spikes, unresolved at 10\,$\mu$s, rising to many times the average pulse peak flux density. Such spikes became known as `giant micropulses': extremely bright spikes that typically do not cause the pulse-integrated fluence to increase by a large factor. The threshold is usually defined as a peak flux density of the spike ten times the average pulse (AP) peak flux density \citep[e.g.,][]{Cairns_2004}. The most famous example of these are the Crab nanopulses \citep{Hankins+2003}. However, already in the early days it was also shown that some pulsars emit so-called ``giant pulses,'' in which the pulse-integrated flux density, i.e., the pulse fluence, is much greater than that of the AP. This was first discovered from the Crab pulsar \citep{1968Sci...162.1481S}, and again the now common definition is that a GP has at least a ten times greater fluence than the AP. By these definitions, the phenomenon is rare: according to the ATNF Pulsar Catalogue\footnote{\url{http://www.atnf.csiro.au/research/pulsar/psrcat/}}, currently there are 2659  known pulsars \citep{2005AJ....129.1993M}, of which only 24 have been observed to produce giant pulses \citep{Kazantsev_2018}.

It has also been shown that the fluence of a pulsar's AP can vary by up to an order of magnitude over the course of a few days. This long term variability in the AP itself is attributed to propagation effects, such as refractive or diffractive interstellar scintillation, RISS and DISS, respectively \citep{1993ApJ...403..183G, 2005MNRAS.358..270W}. However,
both the giant pulses and giant micropulses are single-pulse events, with the next pulse usually being `normal', which implies that they come and go too fast to be caused by any interstellar propagation effect \citep[e.g.,][]{2012AJ....144..155S}. The factor-10 thresholds for calling something a giant pulse or giant micropulse appear to be arbitrary. In well-studied cases there is not a bimodal distribution of instantaneous peak fluxes or pulse fluences, as revealed for example in the detailed studies of PSR\,B0950$+$08 itself \citep{Cairns+2004} or in the study by \cite{Kazantsev_2018} of the pulse flux distributions of a number of pulsars in which some new giant-pulse producers were discovered. 
It may even be that both giant pulses and giant micropulses are parts of a continuum of strength and structure in individual pulsar pulses. On the other hand, the giant pulses are generally narrower to much narrower in duration  \citep[e.g.,][]{Hankins1971,1538-3881-151-2-28,Johnston+2001} and have intricate patterns in their frequency structure and in the phases where they appear, as best studied in the Crab pulsar \citep[][and references therein]{Hankins+2016}. Also, there is a strong preference for the GPs to line up in phase with high-energy emission of the same pulsar \citep{JohnstonRomani2002}.

Here we report on the detection of extreme GPs, with fluences up to 400 times that of the AP, from the pulsar PSR\,B0950$+$08 by AARTFAAC. Our time resolution of 1\,s precludes us from studying any details of the pulses, but our very long observing time of almost 100\,h or 1.4 million pulse periods is the longest thus far reported, rivalled only by the study by \cite{2012AJ....144..155S}, who observed for just over 75 hours. We have therefore collected very good statistics on the occurrence of these pulses; their properties are somewhat discontinuous with those reported before, and so we have either found a novel manifestation of GPs, or extended their range of properties. 
AARTFAAC is a radio transient survey project attached to the LOw-Frequency ARray (LOFAR), whose aim is to search for bright (tens of Jy), brief (seconds to minutes timescale), transient events around 60\,MHz \citep{2016JAI.....541008P}. We first found the brightest of these GPs as part of a blind, general survey \citep{Kuiack+2020} and then focused on the location of the pulsar for a deeper search, to unveil a large sample of GPs.

In Section \ref{sec:obs} we describe our instrument, observations, and data processing methodology. We then report details of the observed pulses and sample statistics in Section \ref{sec:results}. Next, in Section \ref{sec:discussion} we discuss evidence for the origin of the extremely bright GPs. Finally, we state our main conclusions in Section~\ref{sec:conclusions}.

\section{Instrument and Observations}
\label{sec:obs}

AARTFAAC is an all-sky radio telescope experiment, designed to survey the Northern Hemisphere for bright low-frequency transient events that evolve on timescales from one second to several minutes \citep{2016JAI.....541008P}. 
It operates as a parallel back-end on  LOFAR \citep{2013A&A...556A...2V}. 
The radio signals  from 288 inverted-V antennae in  the central six low-band antenna (LBA) stations  are split off digitally and then processed by  dedicated correlators and calibration and imaging servers. 
The longest baseline is 300\,m, resulting in an angular resolution of 60'. Since we use the dipoles individually (as opposed to LOFAR's more standard practice of summing station dipoles to a beam), we retain the full dipole field of view, which is the full sky. 
However, due to ground-based radio frequency interference (RFI), and a significant drop in sensitivity towards the horizon, the detection region for transients is set to 50$^\circ$ from Zenith (Declination = $52.9^{\circ}$), yielding a field of view of 4800 square degrees. 
This allows for the detection of $30--50$ persistent sources, per second, across the field of view, with a significance of $5\sigma$ ($\sim50$ Jy during typical conditions). The parameters used were settled upon by trial and error, and seek to balance the number of candidate sources while reducing noise \citep{10.1093/mnras/sty2810}.

\begin{figure}
\includegraphics[width=\columnwidth] {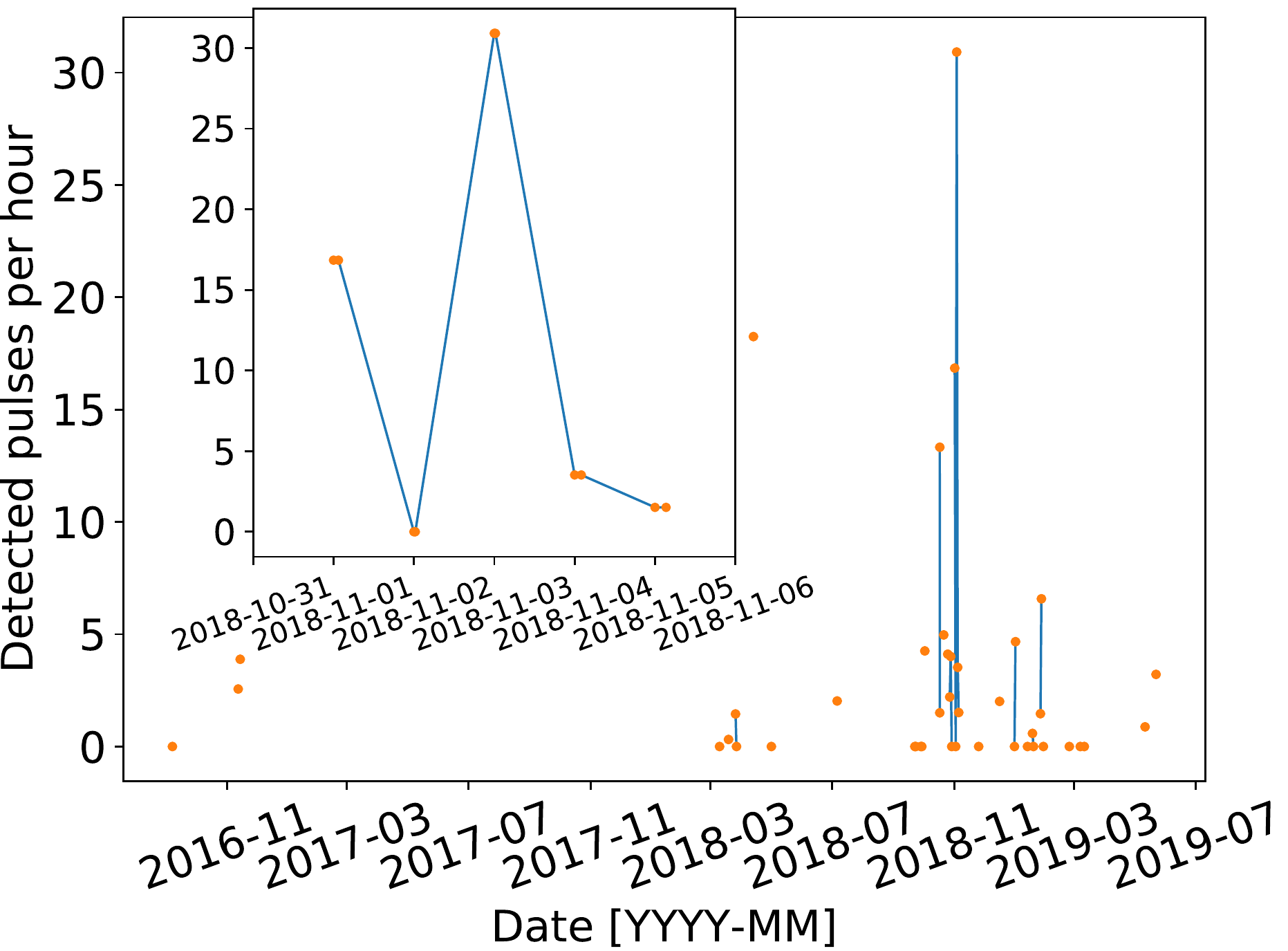}
\caption{The rate of pulses detected over the course of our observations. The observation start time is marked with an orange dot; observations within 24 hours are connected with a blue line. The rate varies greatly from day to day.  The insert shows the high day to day variation in the rate of detected pulses.    } 
\label{fig:obs_time-corr}
\end{figure}

We observe simultaneously with any regularly scheduled LOFAR-LBA observations, or during ``filler time'' when no observation has been scheduled. 
Therefore, the archived data sets vary in length and are spaced irregularly in time, with LST-matched observations over consecutive days occurring rarely in the schedule. 
Although LBA and filler time observations represent a minority of the total scheduled LOFAR observations, the AARTFAAC archive has accumulated  over 1100 hours of observations between August 2016 and September 2019. 
The complete set of observations used in this work consists of 96\,h between 2016-09-07 and 2019-05-21 in which PSR\,B0950$+$08 was within the search area.
The dates of these observations, as well as the rate of pulses detected per hour, are illustrated in Fig.~\ref{fig:obs_time-corr}.

The data were recorded as 16 separate, 195.3 kHz bandwidth, calibrated subbands, with an integration time of 1\,s. 
Visibility calibration is described fully in \mbox{\cite{2014A&A...568A..48P}}, but put simply, the brightest sources (the A team and the Sun) are subtracted with single component Gaussian models, and the diffuse Galactic emission is reduced by imposing a minimum baseline length of $10\lambda$. 

The blind search for radio transients is done by first generating two all-sky images, $1024\times1024$ pixels, each second, with a bandwidth of 1.5\,MHz, centered at 58.3 and 61.8\,MHz, near the peak of the LBA sensitivity. 
The images are then processed by the LOFAR Transient Pipeline \citep[TraP;][and references therein]{2015A&C....11...25S}\footnote{\url{https://github.com/transientskp/tkp}}, which ingests an image set from an AARTFAAC observation, performs source finding in each image, associates sources across images, while calculating detection significance and variability statistics. 

Most of the initially detected sources are spurious, and thus we set very strict thresholds for detection and filtering \citep[for full details, see][]{Kuiack+2020}. It helps to retain two 8-subband images,
despite the sum of the two having better signal-to-noise,
since many spurious signals are narrow-band or strongly frequency dependent and can thus be rejected by comparing the two images. 
For the blind search, a viable source must be detected at $8\sigma$ or more in one of the two images, and simultaneously at $5\sigma$ or more in the other.

The flux scale is computed for each image individually by using a linear least squares fit of the instantaneous brightnesses of all detected sources in the AARTFAAC catalogue \citep{10.1093/mnras/sty2810} to their catalogued flux values.  
The validity of the flux scaling is measured by comparing the extracted flux densities of the AARTFAAC catalogue sources to their catalogued value.
There is no appreciable systematic offset (<1\%), and the average uncertainty is 30\%, which is the normal value due to measurement error and ionospheric variability.

\begin{table*}
\begin{center}
\caption{Comparison of the average pulse (AP) and giant pulse (GP) parameters, including pulse fluence distribution power-law index, across the different observations.  
$\dagger$ values are calculated using linear interpolation of the other parameters.}\begin{tabular}{ r@{}l r@{}l r@{}l r@{}l r@{}l l }
\hline\hline
\multicolumn{2}{l}{Freq.}  & \multicolumn{2}{l}{Period-averaged} & \multicolumn{2}{l}{AP fluence} &  
     \multicolumn{2}{l}{Max GP} & \multicolumn{2}{l}{Index} & Source \\ 
\multicolumn{2}{l}{[MHz]}  & \multicolumn{2}{l}{flux [Jy]}       & \multicolumn{2}{l}{Jy\,s}      &     
     \multicolumn{2}{l}{fluence [Jy\,s]} & &  \\ \hline
112&     & 2&            & 0&.51          &  81&.2  & $-1$&$.5$       &  \cite{2012ARep...56..430S} \\ 
103&     & 3&            & 0&.76          & 228&    & $-2$&$.6\pm0.2$ &  \cite{2012AJ....144..155S} \\ 
74&      & 2&.4          & 0&.61          &  14&.6  & $-5$&$.06$      & \cite{1538-3881-151-2-28} \\ 
61&.8    & 2&.6$\dagger$ & 0&.66$\dagger$ & 155&    & $-4$&$.8\pm0.2$      & This work\\ 
58&.3    & 2&.6$\dagger$ & 0&.66$\dagger$ & 177&    & $-4$&$.5\pm0.1$     & This work\\ 
42&      & 2&.8          & 0&.71          &  17&.0  & $-4$&$.09$      & \cite{1538-3881-151-2-28} \\  
39&      & 1&.5          & 0&.38          &  10&.7  & $-4$&$.7$       & \cite{1538-3881-149-2-65}  \\
\hline\hline
\label{tab:GPcompare}
\end{tabular}
\end{center}
\end{table*}

\section{Results}
\label{sec:results}

\begin{figure}
\includegraphics[width=\columnwidth] {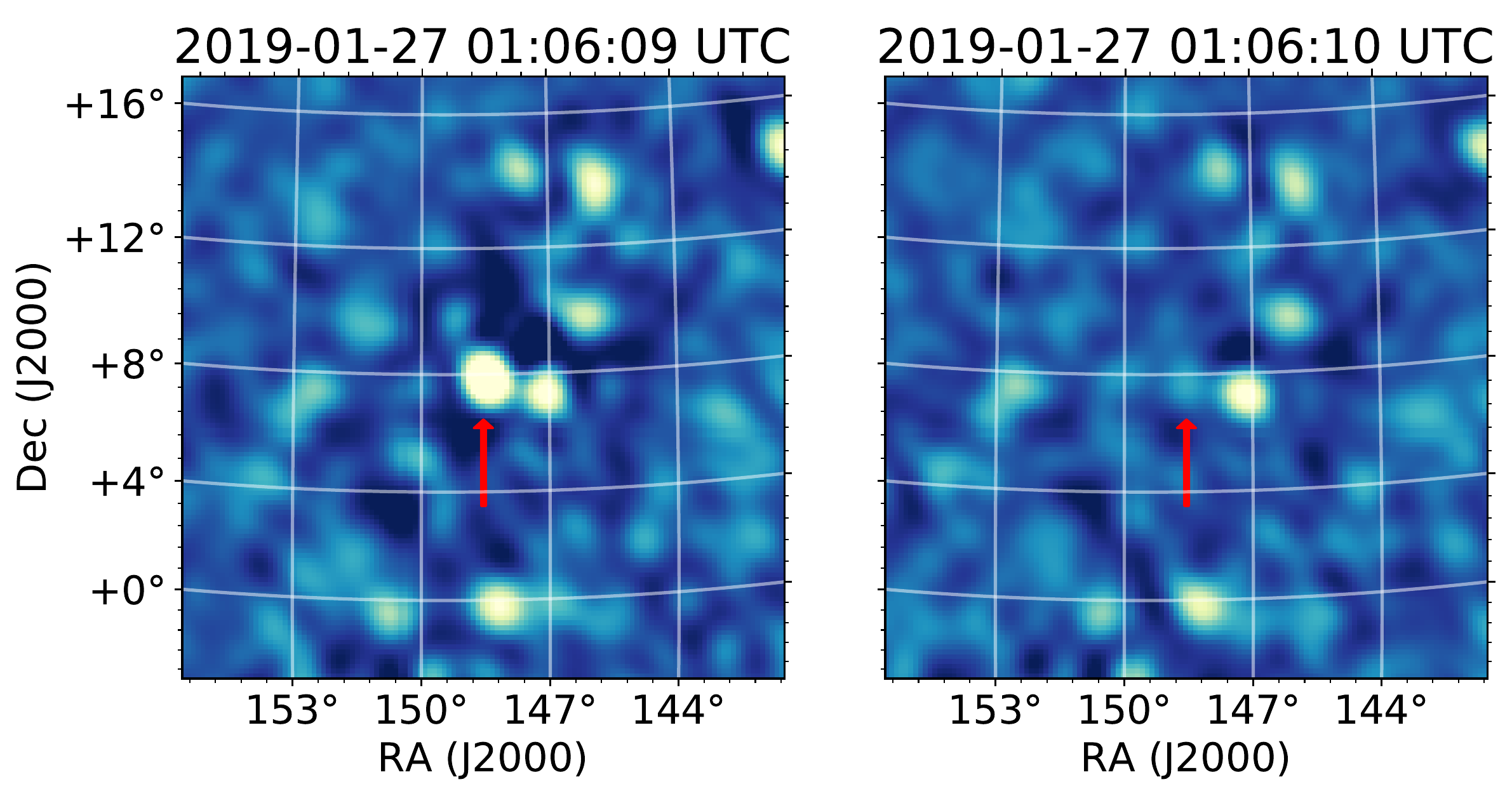}
\caption{Example image stamps showing one of the brightest pulses detected, with a fluence of $177 \pm 53$\,Jy\,s (left), while the  background in the next frame is $34 \pm 25$\,Jy\,s (right); these fluences are the integrated flux density over the 1\,s exposure time of the image. The FWHM of the AARTFAAC PSF is 60'. The source immediately adjacent is 3C\,227, a Seyfert 1 galaxy with a flux density of $113.7 \pm 0.1$\,Jy at 60\,MHz, in the AARTFAAC catalogue.  } 
\label{fig:examplepulse}
\end{figure}

The signal from PSR\,B0950$+$08 was originally discovered blindly as a transient candidate when analyzing the transient survey observation from 2018-11-01. Once we knew this source was producing transient events, we reanalysed the data at its location, lowering our detection threshold from 8 to 5$\sigma$, because the number of random false positives goes down very much when searching only one pixel per image rather than one million.
Additionally, observations stored in the visibility archive in which PSR\,B0950+08 was within the field of view were prioritized for processing until a sufficient sample of pulses were detected. 
In total 275 pulses were detected in 96 hours of observations (Fig.~\ref{fig:obs_time-corr}); given the low declination of the pulsar, we never get more than 4\,h of consecutive data from within our search region ($Z<50^\circ$).
In the inset of the plot we show the rate of GPs detected per hour during a week in early November 2018. Clearly very strong day-to-day variations in the rate are possible. 
The brightest pulse detected, with a fluence of $177 \pm 53$\,Jy\,s, is shown in Fig.~\ref{fig:examplepulse}.

\begin{figure}
\includegraphics[width=\columnwidth] {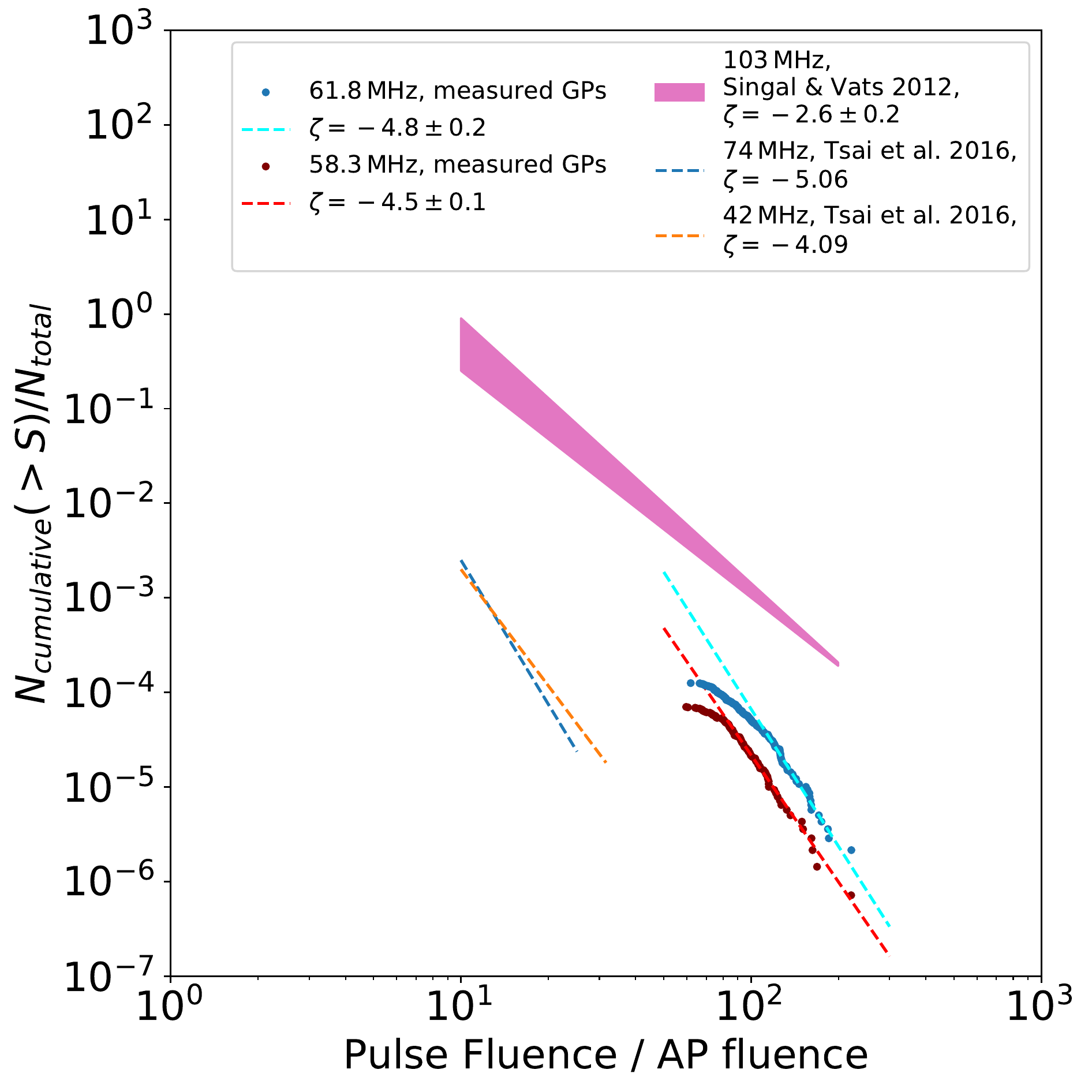}
\caption{Cumulative fluence distributions from all of the pulses detected thus far in the AARTFAAC survey (red and blue dots), compared to estimations of the pulse fluence distributions fit at 74 and 42\,MHz observed by LWA (dark blue and orange dashed lines), and at 103\,MHz using the Rajkot radio telescope (purple). The power-law distributions are fit to pulses with fluence to AP fluence ratios greater than 150 at 61.8\,MHz (light blue dashed line), and 120 at 58.3\,MHz (red dashed line), to avoid the effect of incompleteness in the sample at low fluences. Clearly the distributions detected by the three experiments are different, providing further evidence of the high degree of variability over time of the GPs of PSR\,B0950$+$08.}
\label{fig:GPbothfreq}
\end{figure}

\subsection{Pulse fluence distribution}

The statistical distribution of all pulses detected thus far shows striking similarity, in power-law index, to the sample of giant pulses observed by \cite{1538-3881-151-2-28} using the Long Wavelength Array (LWA), an instrument similar in design to AARTFAAC. Shown in Fig.~\ref{fig:GPbothfreq} are the power-law indices of the GP fluence distribution at 42 and 74\,MHz as measured by \cite{1538-3881-151-2-28}; they agree well with our results at 58.3 and 61.8\,MHz. (Note that since they observed the GPs in beam-formed mode with high time resolution, comparing the fluences is non-trivial, see Sect.~\ref{sec:discuss:flu}.)
However, the rate at which we are observing pulses is about five orders of magnitude too high, given that our pulses are an order of magnitude brighter, if we try to extrapolate both distributions to a common fluence. The much higher rate at which we observe GPs indicates that this is clearly not merely the high-fluence tail of the previously observed distribution. However, \cite{1538-3881-149-2-65,1538-3881-151-2-28} observed for 36\,h within a short time period, so a very low point in the GP rate may have contributed to the difference. If we compare with the much longer-term study of \cite{2012AJ....144..155S}, who observed 141 30-min intervals over a period of ten months we see a better agreement in rate, but a different slope: the distribution of GP fluences they measure during the most active periods has a shallower slope and lower mean fluence, but an even higher GP rate on their most active days than we find on our most active days in the overlap (Fig.~\ref{fig:GPbothfreq}).

Furthermore, extrapolating the distribution fit by \cite{1538-3881-151-2-28}, to the pulse to AP fluence ratio of 1 yields a $N_{\mathrm{cumulative}}(>S) / N_{\mathrm{total}} \approx 1$. This indicates that the pulses reported there could be drawn from a single distribution which extends continuously all the way to the normal pulse fluences, similar to the cases reported by \cite{Kazantsev_2018}. By contrast, extrapolating the pulse fluence distribution observed by AARTFAAC to the AP fluence is not possible, it would require far more GPs than there are pulse periods in our data set. The same is true for the data from Rajkot \citep{2012AJ....144..155S}, although a bit less so due to the shallower slope. This motivates us to see whether we can also see evidence for a shallower slope at lower fluences or a dependence of the slope on the GP rate. 

\begin{figure}
\includegraphics[width=\columnwidth] {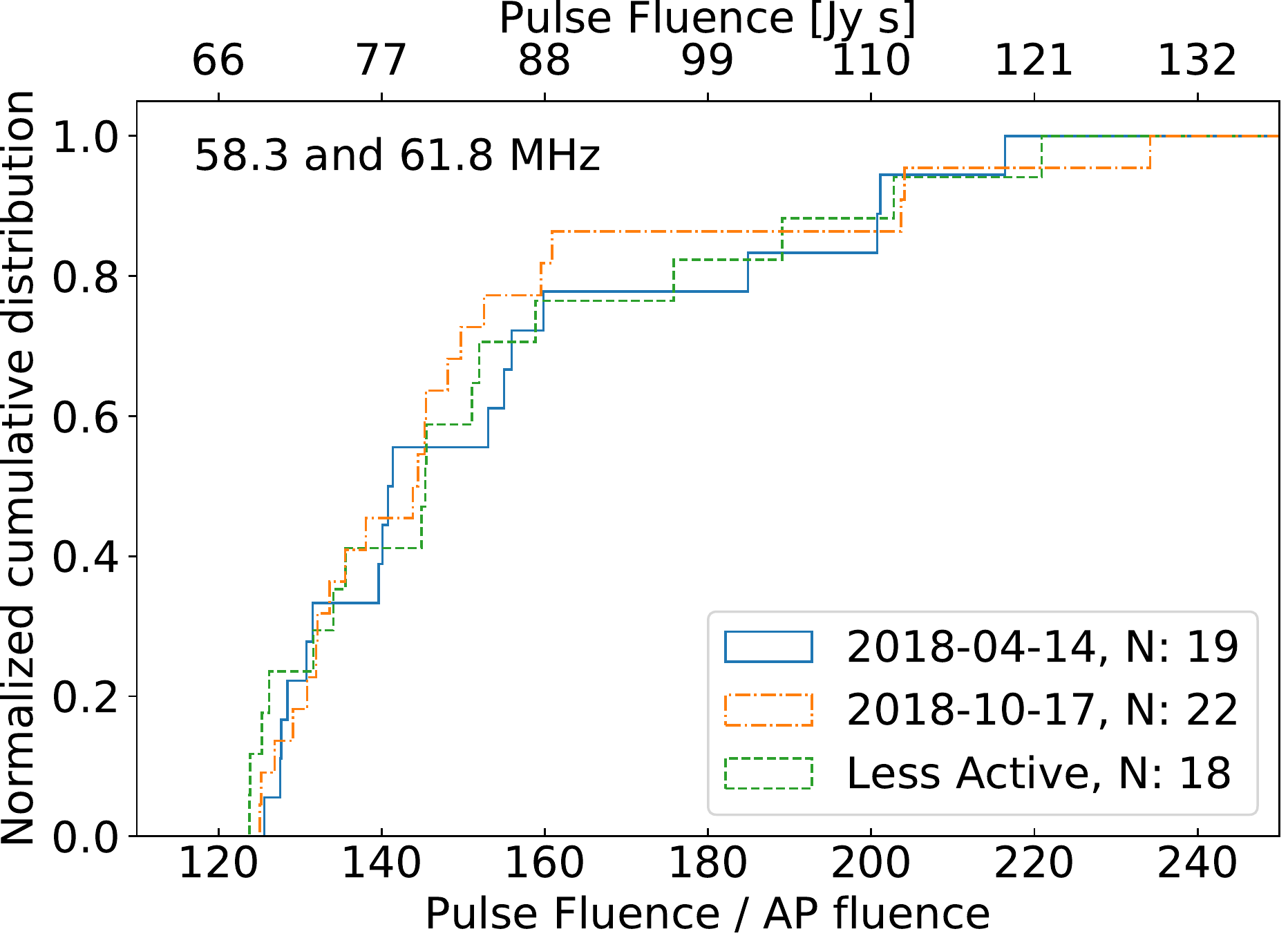}
\caption{The cumulative fluence distributions of pulses from the two observations with the highest activity as well as a collected sample of pulses from observations when the pulsar appears less active. }
\label{fig:KStest}
\end{figure}

To start with the latter, we compare the fluence distributions of three periods: the two most active days, 2018-04-14 and 2018-10-17, as well as a collection of pulses from the less active days. We compare them using two-sample KS tests, illustrated in Fig. \ref{fig:KStest}. 
In order to ensure that variations in the sensitivity across multiple observations are not a confounding factor in the shape of the distributions, first a completeness limit was applied. 
We only compare pulses with a fluence greater than 68\,Jy\,s, occurring when the detection limit was less than 68\,Jy\,s. 
Fig. \ref{fig:KStest} shows the similarity of the three samples. 
This is made more quantitative using two-sample KS tests: 
the null hypothesis that the samples are drawn from the same distribution is not rejected when comparing any two of the three distributions. But of course the samples are small, so only large variations in the slope are excluded by this test. In the next subsection we describe how we extended the study to fainter GPs.

\subsubsection{Lower-fluence pulses and their distribution}
\label{sec:forcefit}

\begin{figure*}
\centering
\mbox{
{\includegraphics[width=0.5\textwidth]{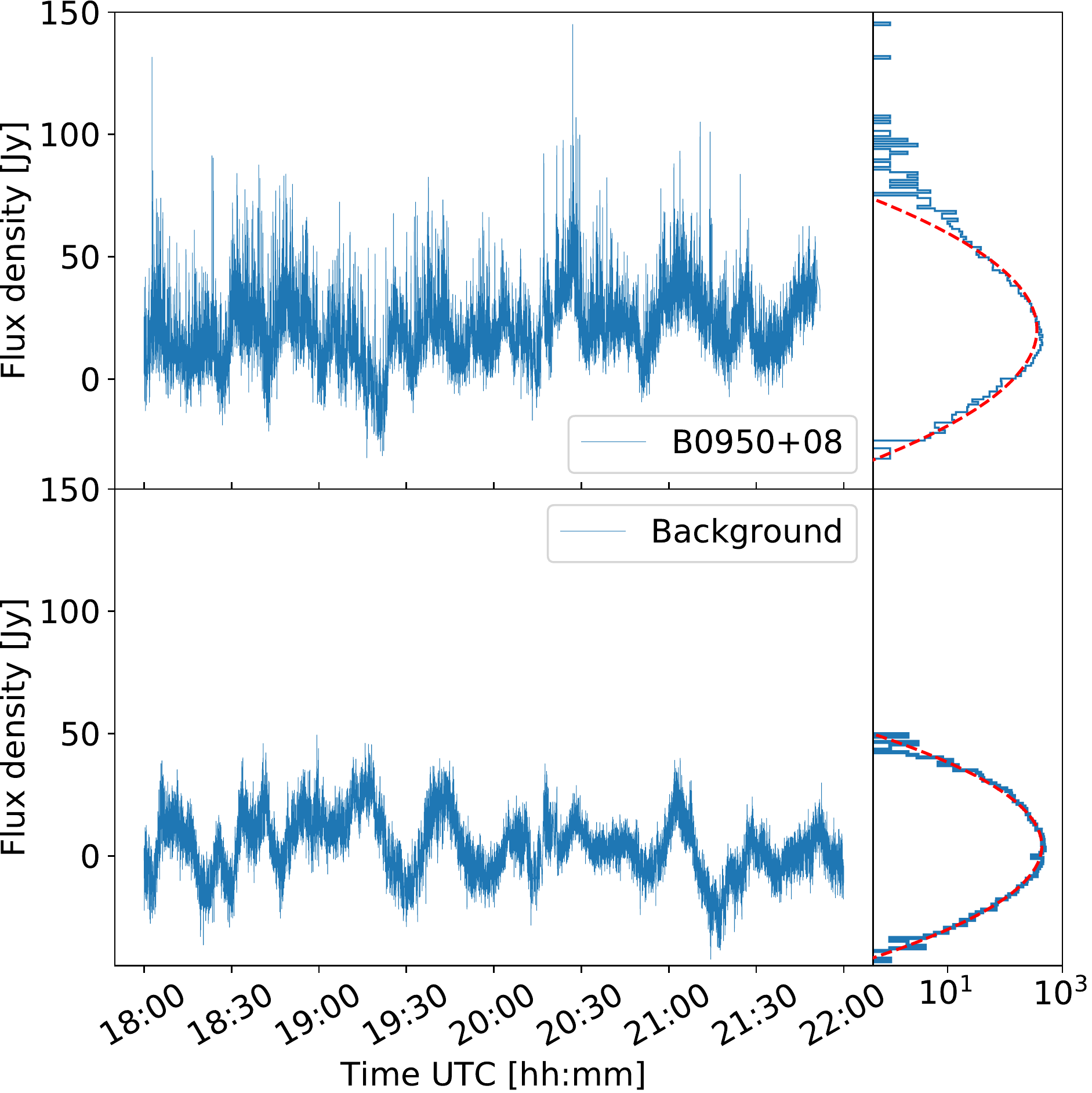}}
{\includegraphics[width=0.5\textwidth]{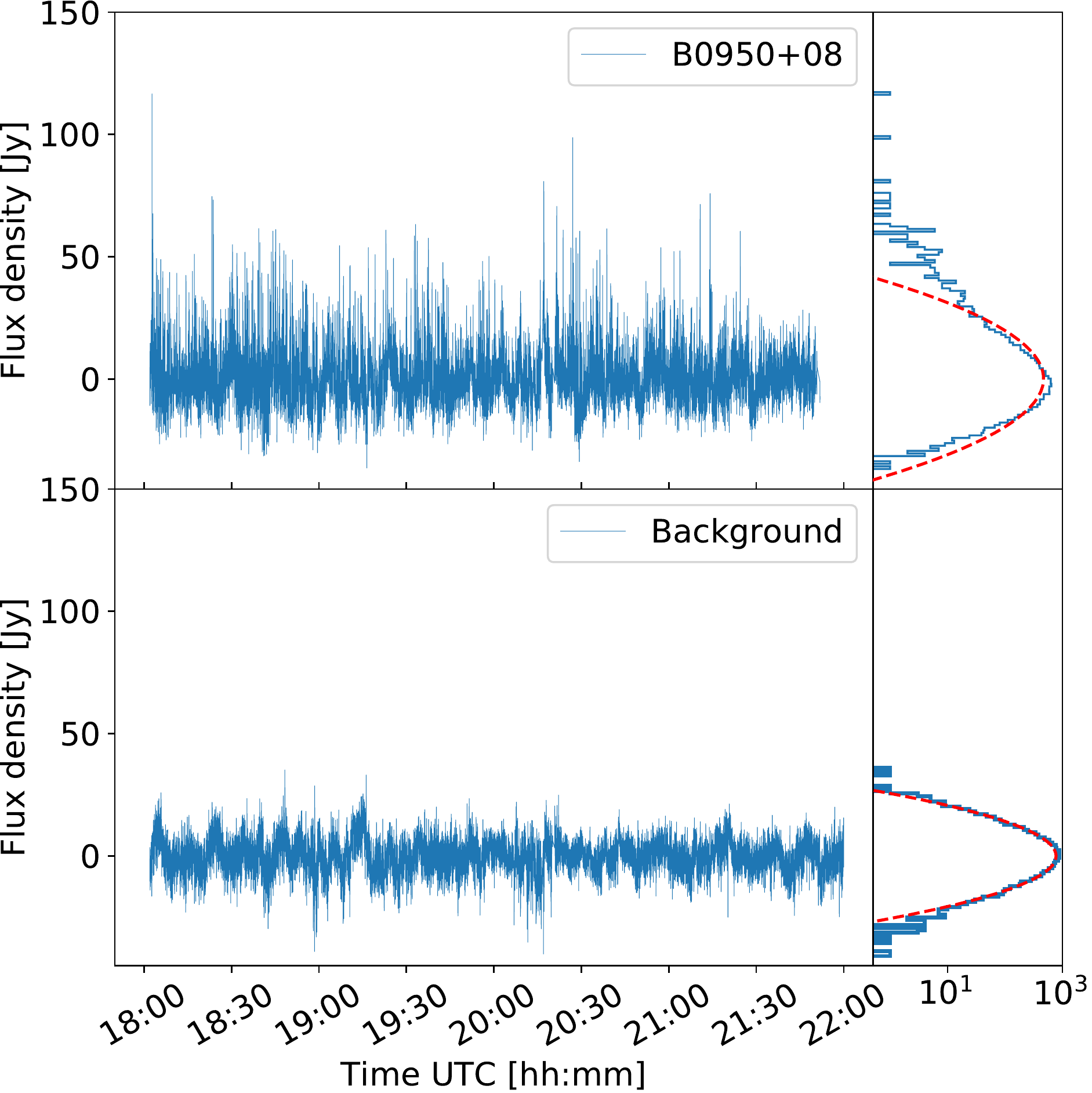}}}
\caption{Fully sampled light-curve of PSR\,B0950+08 (top) and a nearby background location (bottom). Both raw light curves  (left) show correlated noise variations which have been removed with a 2\,min  boxcar rolling mean (right). The red dashed limes show a Gaussian fit to the flux density measurements from each light curve. These fits show how well the background is modelled by a Gaussian, whereas the pulsar shows an obvious over abundance of high flux density measurements. }
\label{fig:fulllightcurves}
\end{figure*}

A significant cause of the high detection threshold applied so far is the fact that the $\sim$30\% ionospheric variations of the source fluxes contribute to the rms.\ we measure, and therefore raise the detection threshold. However, as noted above, these variations occur on a rather slow time scale of 10--15\,min compared to the 1\,s timescale of the GPs. Therefore we can try to establish a lower noise threshold by subtracting the ionospheric variations through low-pass filtering of the light curve. We also do this for some nearby locations where there is no bright source, in order to see the effect of this process on the noise alone. 
The observation containing the most active time period was reprocessed with the TraP, resulting in a fully sampled light curve from 18:00 to 22:00 UTC on 2018-04-14, when PSR\,B0950+08 was within the detection region of the image ($50^{\circ}$ from Zenith). This resulted in the light curves illustrated in Fig.~\ref{fig:fulllightcurves}.
The light curve shown in the bottom left panel gives the forced flux measurements at a location on the sky a few beam widths away from PSR\,B0950+08, where no source is visible to AARTFAAC. We determine the ionospheric variation 
by modelling it with a rolling boxcar mean with a width of 2\,min. This is short enough to remove the longer-term variability, and long enough that GPs do not significantly affect the rolling mean. 
The result of subtracting the boxcar mean is shown in the bottom righthand panel of Fig. \ref{fig:fulllightcurves}. One can indeed see from the histograms attached to the right of each panel that the rms.\ goes down considerably, and that the intensity histogram is very well fit by  a Gaussian after the filtering.
In the top panels we show the result of the same procedure applied to the location of PSR\,B0950+08. Its background-subtracted light curve (top right panel of Fig.~\ref{fig:fulllightcurves}) now clearly shows a great many more GPs, and the intensity histogram shows a Gaussian with the same width as the background panel below, plus a highly skewed tail due to the many more GPs than previously detected.

\begin{figure*}
\includegraphics[width=\textwidth] {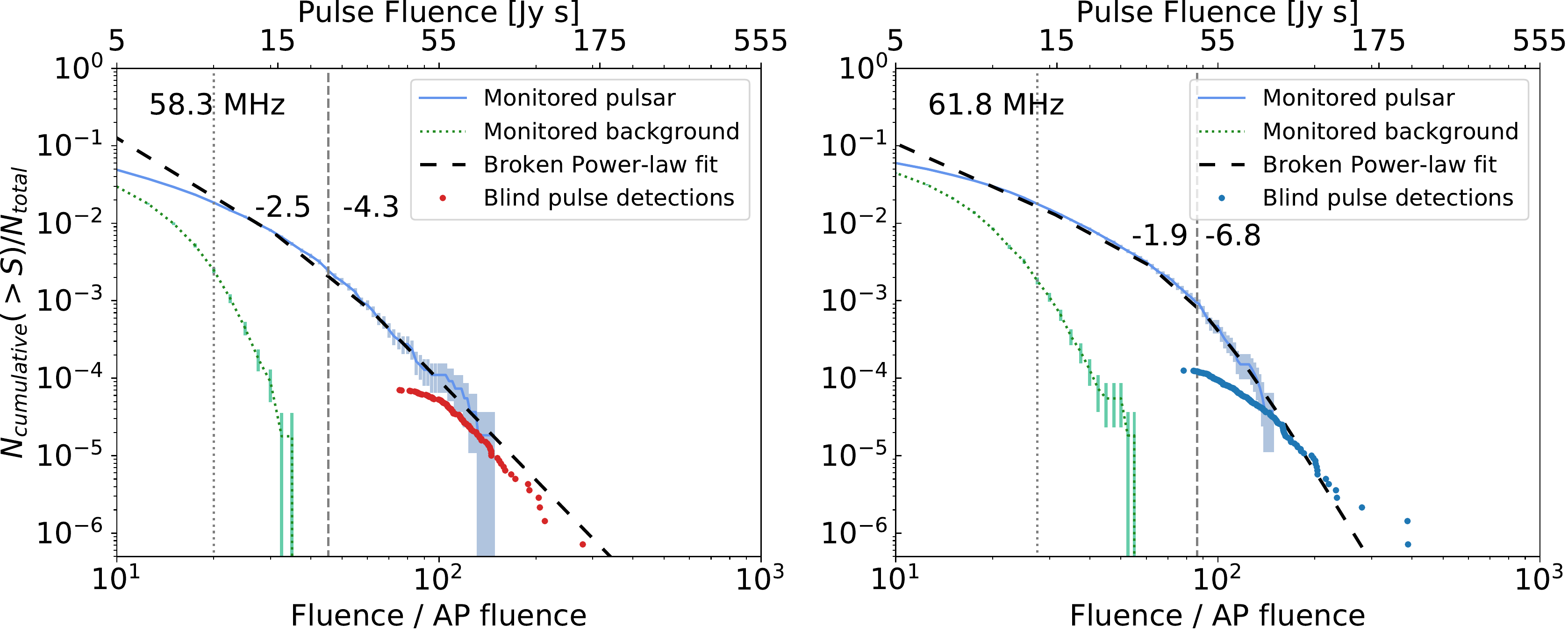}
\caption{Extending the distribution of GPs to lower fluences using the monitoring measurements from Fig. \ref{fig:fulllightcurves}, for 58.3\,MHz (left) and 61.8\,MHz (right). As in Fig. \ref{fig:GPbothfreq}, the horizontal axis is normalised to the AP fluence, and the vertical axis to the total number of rotation periods in the measured time interval. The dots show the pulses detected in the blind search, copied from Fig.~\ref{fig:GPbothfreq}. The pulse fluence distribution from the lower-threshold search is shown in light blue (with Poisson error bars) and the background fluence distribution at the same time is shown in green (again with Poisson error bars). The black dashed line shows a broken power-law to the overall pulse fluence distribution, with the grey vertical dashed line indicating the location of the break, and the numbers to either side of it the asymptotic low- and high-fluence power-law indices. The grey dotted line shows the fluence above which the background contributes less than 10\% to the distribution.}
\label{fig:bgsubcumulative}
\end{figure*}

In figure~\ref{fig:bgsubcumulative} we show the depth to which AARTFAAC is able to probe the fluence distribution of PSR\,B0950+08 emission after this careful accounting of the background noise.
In each panel we depict the sample of GPs detected in the transient candidate database via the blind search criteria, compared to the distribution of flux measurements made every second at the location of the pulsar and the nearby background. 
We find that we can reach about 5 times lower fluences and can indeed see a break to a shallower slope at low fluence. Detailed parameters of a smoothly broken power law are not very well constrained, but we show illustrative fits to the data that indicate a break around a fluence to AP fluence ratio of about 45 (86)
at 58.3\,MHz (61.8\,MHz), and slopes below the break that are comparable to those measured by \cite{2012AJ....144..155S} at 103\,MHz.

\subsection{Pulse clustering}

\begin{figure}
\centering
\includegraphics[width=\columnwidth]{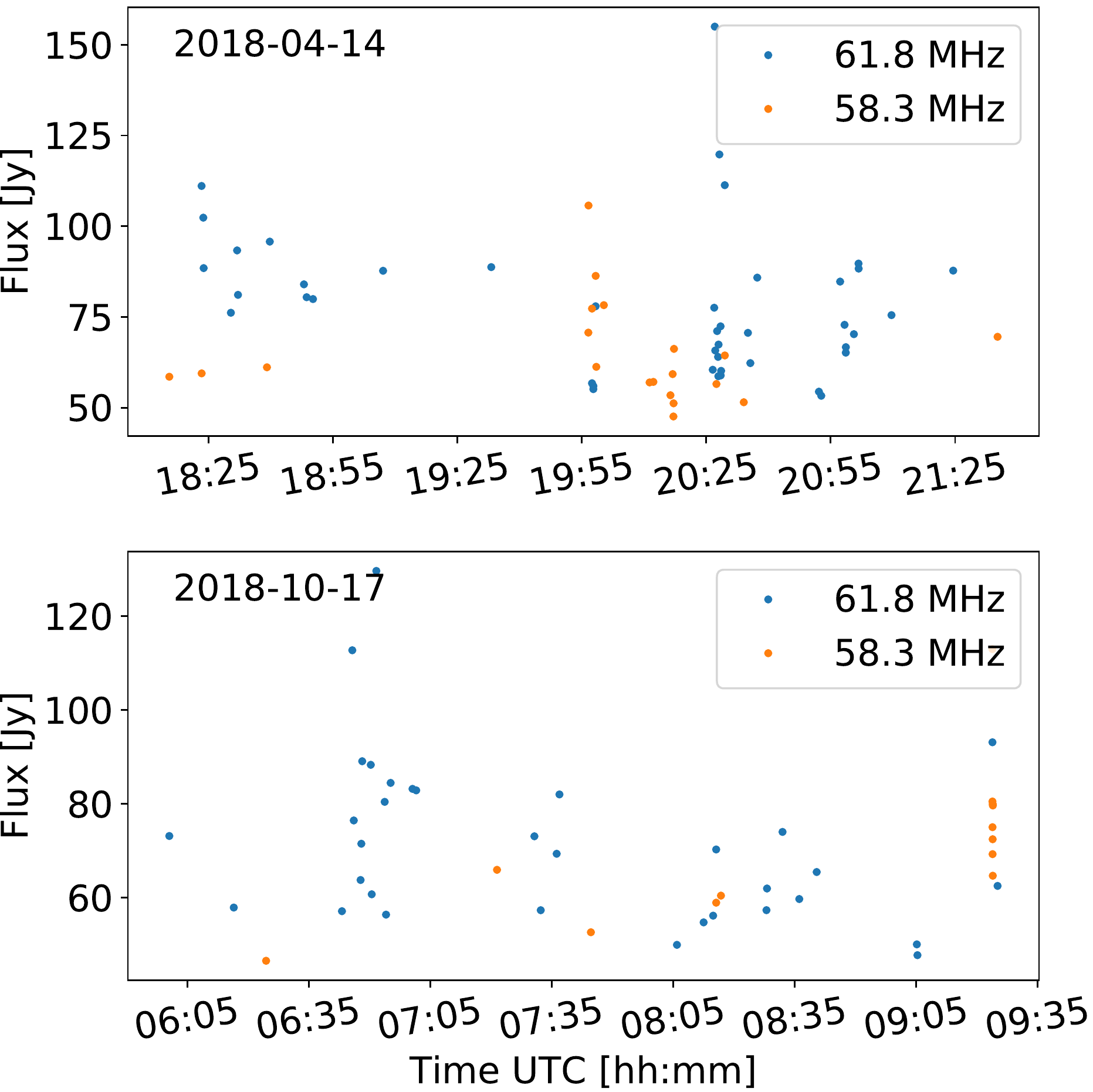}
\caption{Pulse arrival times, coloured by the detection frequency from the two most active days. The dispersion index measured for a range of bin time bin widths is greater than 1, indicating that the pulse behaviour is clustered more than expected from a Poisson process.} 
\label{fig:pulsetime}
\end{figure}

\begin{figure}
\centering
\includegraphics[width=\columnwidth]{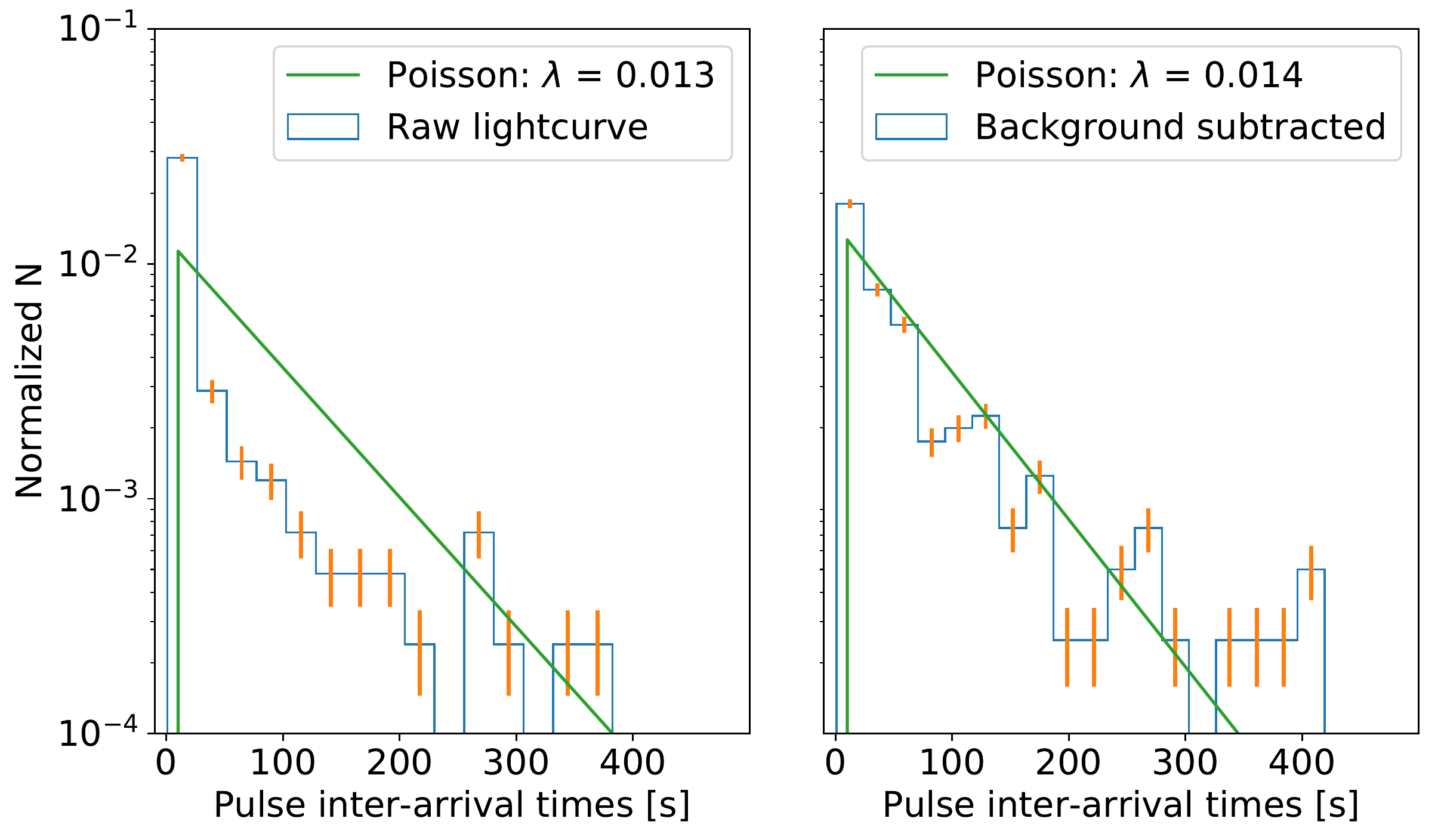}
\caption{ Histograms of the inter-arrival times for pulses above fluence 50\,Jy\,s. Counting uncertainty is the square root of the number of counts. The left panel shows the distribution for the raw light curve, and the right for the background subtracted light curve. The raw light curve is clustered, whereas once the background is removed, the distribution is exponential decay, consistent with a Poisson process.} 
\label{fig:clusterdist}
\end{figure}

To investigate our impression that the GPs vary greatly not only between days, but also come in clusters within a given observation, we perform some formal tests of clustering on the GP time series within the two most active days.
The clustering of time series can be measured in two ways, both by comparing to what would be expected from a Poisson process. First we can compute the so-called dispersion index, the ratio of the variance to the mean of binned arrival times. A dispersion index of 1 is expected for a Poisson process, less than 1 indicates overly regular data, and greater than 1 indicates clustered data. This has the drawback of being dependent on the width of the bins used, but for all reasonably sized bins (30\,s to 10\,min) the dispersion index is  greater than 1 during the two most active days (Fig.~\ref{fig:pulsetime}), indicating a greater tendency for giant pulses to appear in clusters. Secondly, the clustering can be measured by comparing the pulse inter-arrival times, the interval of time between successive GPs. For a Poisson process the distribution of inter-arrival times should be exponential. For both days, the distribution of inter-arrival times was not well fit by an exponential distribution; there was a greater number of short inter-arrival times, illustrated in the lefthand panel of Fig.~\ref{fig:clusterdist}. 

However, as can be seen in the top left panel  of Fig.~\ref{fig:fulllightcurves}, where we show the full light curve for 2018-04-14, the brightest GPs follow the modulation due to the ionosphere, and this may contribute to the detected clustering. In order to examine this we re-perform the Poissonian test on the data with the background modulation filtered out. 
This completely eliminates the clustering: 
the corrected distribution of pulse inter-arrival times is shown in the right hand panel of Fig.~\ref{fig:clusterdist}, and follows an exponential distribution, in good agreement with Poisson statistics. Since the day-to-day mean sensitivity variations are much smaller, the large variations in daily rates cannot be explained in the same way as due to sensitivity, and so we conclude that the GP rate of PSR\,B0950$+$08 is highly variable on all timescales from days to years, but not on a timescale of hours.

\subsection{Pulse spectral behaviour}

\begin{figure}
\centering
\includegraphics[width=\columnwidth]{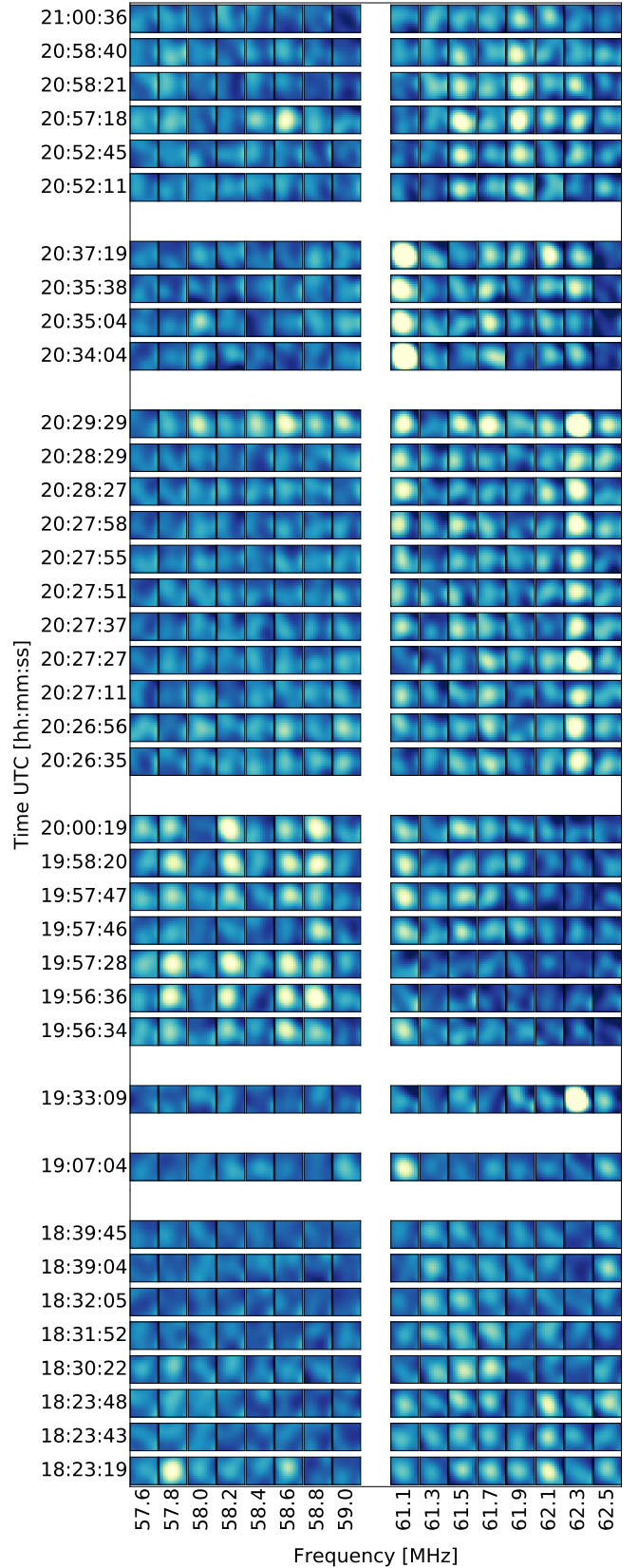}
\caption{ Time/frequency plot of the brightest pulses from 2.5 hours on 2018-04-14. It is clear that the spectral structure is not the result of DISS, which has a bandwidth of $\approx65$ kHz and would average out over only a few of the 195.3 kHz subbands. } 
\label{fig:freqtime}
\end{figure}

In order to inspect the spectral structure of the GPs, we re-imaged all 195.3\,kHz subbands separately.  The observation used was recorded on 2018-04-14; this date was chosen because it contained the largest number of detected pulses. All of the pulses detected in the blind search mode are plotted in the top panel of Fig. \ref{fig:pulsetime}. 40 of the brightest pulses spanning 18:23 to 21:01 UTC are illustrated in Fig. \ref{fig:freqtime}. The image stamps have a consistent colour scaling, and provide a qualitative comparison of the relative flux in each frequency channel. The stamps are separated vertically according the clusters of time, and horizontally in the two consecutive sequences of subbands.  
Interestingly, we find that the spectra of the pulses are highly structured and variable, with many narrow-band pulses, and with pulses close in time tending to have similar structure.

\section{Discussion}
\label{sec:discussion}

\subsection{Comparing GP fluences properly}
\label{sec:discuss:flu}

Because our integration time is four pulse periods long, one might worry that our high fluences are caused by the fact that we sum the fluence of multiple pulses. This does seem unlikely, however, given that the GPs are still relatively rare (about 1 per 10,000 pulses for the brightest blindly detected ones). So two GPs occurring in one second by random chance will be very rare, and also there is no prior evidence that GPs are strongly clustered on the shortest time scales. We can also call in the help of dispersion:
the DM for PSR\,B0950+08 is $3.0\,\mathrm{~pc~cm^{-3}}$, which
leads to a dispersion delay between the two AARTFAAC images, centered at 58.3 and 61.8 MHz, of 0.4\,s. Since the
rotation period is $0.253$\,s and the pulse duration only 25\,ms, dispersion should move the flux of a GP completely from one time bin to the next if the fluence increase is due to a single GP, and only partly if the fluence increase is often due to multiple GPs in one time bin. 
Pulse signals arriving in the higher-frequency image before the lower frequency image have been observed in a number of cases, and the split is always complete, so we can be confident that the majority of measured GPs are individual pulses. 

Because some observations have resolved the individual pulses in time and report the flux densities, whereas others --in fact most-- do not resolve the GPs, we must be careful to compare all the results to ours on a common scale. For some, such as the long series by \cite{2012AJ....144..155S}, this was fairly easy because even though they report their results in flux density, they already scale the GPs to the AP, and because they resolve neither this is equivalent to scaled fluences. The same holds for the temporally unresolved giant pulses at 112 MHz \citep{2012ARep...56..430S}, and AARTFAAC.
The LWA papers \citep{1538-3881-149-2-65, 1538-3881-151-2-28}, to which we also compare in Table \ref{tab:GPcompare} and Fig.~\ref{fig:GPbothfreq}, resolve the pulse profiles of the GPs. 
We therefore have to translate their flux values to fluences for comparison.
The maximum GP fluences, period-averaged flux and GP fluence distribution power-law indices, are given for comparison in Table~\ref{tab:GPcompare}.

\subsection{Cause of widely differing GP fluence distributions}

Since the brightness also of the AP of PSR\,B0950$+$08 is known to vary by about a factor 10 \citep{1968Natur.218..126P,1993ApJ...403..183G,2012AJ....144..155S}
over time scales that are in good agreement with the cause being refractive interstellar scintillation, we should address whether possibly some of what is seen here could be caused by magnification and propagation effects rather than intrinsic variability of the pulsar.

\subsubsection{Magnification due to interstellar propagation}

Without being able to measure the AP during each observation, we cannot determine the ratio between the GPs we observe and the AP from one day to the next. However, both our own observations and those of \cite{2012AJ....144..155S} show that the fluence distribution of GPs changes slope to fainter values, and thus we should expect that if the rate varies due to the entire distribution being magnified, carrying more GPs across our detection threshold, we should see the slope of the fluence distribution flatten from about $-5$ to $-2.5$ when the rates are highest. However, when we tested this we saw no change of the shape between epochs of low and high rates (Fig.~\ref{fig:KStest}) and so we can exclude interstellar magnification as a significant factor. The fact that the GP rate can change strongly between consecutive days, which is too fast for RISS, points to that same conclusion, as does the fact that \cite{1538-3881-149-2-65, 1538-3881-151-2-28} observed a steep distribution slope at much lower fluences.
Of course, a direct way to test this will be to observe simultaneously with a much more sensitive instrument that can also detect the AP and measure its flux directly; it will be difficult to get the large amount of monitoring time required on such a large telescope, but one could trigger a sensitive observation based on the alert of a high GP rate from AARTFAAC. 

In addition to RISS, the pulsar signal is also diffracted by the interstellar medium during propagation (DISS), which causes variations on much shorter timescales and over possibly narrow bandwidths. 
\cite{10.1093/mnras/stw1293} recently measured a DISS bandwidth of 4.1\,MHz and timescale of 28.8 minutes at 154\,MHz using the MWA. 
These values agree well with a previous study at 50\,MHz by \cite{1992Natur.360..137P}, indicating the dynamics of the interstellar medium have been stable over the last three decades. 
This means we can use the fairly common, though not universal  scaling relations for DISS bandwidth $\Delta \nu_{d} \propto \nu^{4.4}$, and timescale  $\Delta \tau_{d} \propto \nu^{1.2}$.
At 60\,MHz, this predicts a bandwidth of 65\,kHz and timescale of 10.3 minutes. Also \cite{2006ARep...50..915S} measure a decorrelation bandwidth of 200\,kHz at 111\,MHz for PSR\,B0950$+$08, which extrapolates to 15\,kHz at 60\,MHz; even with the somewhat shallower scaling suggested by a detailed study of the scintillation in the sightline by \cite{2014ApJ...786..115S}, we still get to only 32\,kHz width.
This timescale agrees fairly well with the timescale over which we see the spectral energy distribution across the 16 AARTFAAC frequency bands change on 2018 April 14 (Fig.~\ref{fig:freqtime}).

However, the spectral structures we see
mostly have a rather larger width of a fair fraction of a MHz, and only occasionally are as narrow as 200\,kHz, so this fits the DISS picture rather less well. 
This makes it more likely to us that the structures present in Fig.~\ref{fig:freqtime} are intrinsic, rather than DISS scintels, but it is hard to be completely sure of this.  

\subsubsection{Intrinsic emission}

Since we conclude that the majority of the effects we observe, and differences with other studies, are intrinsic to PSR\,B0950$+$08, we should compare it with the behaviour of other known GP pulsars. 

First, we look at other examples of GP fluence distributions.
\cite{2013MNRAS.430.2815Z} show a break in the power law between the cumulative probability distributions of GPs from B1937+21. The index changes  from $-1.6 \pm 0.1$ at lower flux levels to $-2.4 \pm 0.1$ at the higher flux tail. 
This is similar to the behaviour observed by \cite{2007A&A...470.1003P} who also show that giant pulses from the Crab pulsar have a break in the power law, the higher flux tail being steeper.
Oddly, the opposite was observed in the distribution of giant pulses observed from the same pulsar by \cite{2019MNRAS.483.4784M} where the distribution flattens significantly, from $-3.48 \pm 0.04$ to $-2.10 \pm 0.02$, at the high flux tail.
When searching for giant pulses, \cite{Kazantsev_2018} found that the typical pulse energy distribution for B0329+54 was clearly best fit by a broken power law, despite not detecting any giant pulses from this pulsar. It was the only pulsar monitored which exhibited this behaviour.

Next, we look at the structure in spectra of GPs and other pulsar emission.
The frequency structure of the individual pulses shown in Fig.~\ref{fig:freqtime} is similar in character to the single pulse spectra shown in some other studies of GPs from different pulsars, at different frequencies.  
This sparse comb-like structure is observed from B1957$+$20 between 338 and 324\,MHz \citep{2017ApJ...840L..15M}, B1821$-$24A between 720 and 2400\,MHz \citep{2015ApJ...803...83B}, and from B1937+21 between 1332 and 1450 MHz \citep{2019MNRAS.483.4784M}. \cite{2012AJ....144..155S} also reported that during a simultaneous observation of PSR\,B0950$+$08 between Rajkot (103\,MHz) and WSRT (297\,MHz) they did not see the same GPs in the different bands simultaneously, and even noted decorrelation within the 10\,MHz WSRT band. \cite{PopovStappers2003} similarly report that for a 3\,h simultaneous observation of  B1937$+$21 at 1.4 and 2.3\,GHz, GPs were found in both bands, but none were simultaneous between them. Also the Crab pulsar GPs \citep[e.g.,][]{Hankins+2016} have very rich frequency structure.
Whether this kind of structure is also present in normal pulses is not well known, since these are usually too faint to make individual spectra of. The APs are spectrally smooth, but of course they are by definition averaged over many individual pulses. 

Lastly, the recent radio outburst from the magnetar RXTE\,J1810$-$197, allowed for the spectral properties to be studied \citep{Maan+2019}. The bursts exhibited high spectral modulation on frequency scales much greater than the scintillation bandwidth, which was then observed to weaken and disappear in the months after the initial outburst. Given the very different nature of this neutron star, a more detailed comparison with PSR\,B0950$+$08 using more sensitive instruments could be very interesting.

\subsection{Pulse rate discrepancy}

Lastly, we discuss highly variable GP rate between the different studies in more detail. 
Of the 96 hours, or 42 observations, which we have used to search for detectable pulses, 18 observations, totalling 37 hours, contained no detectable pulses. 
Assuming the activity level is typical this indicates that roughly $40\%$ of observations could yield no pulses at this extreme fluence level. 
However, in 21 of our observations the rate of detected pulses was greater than one per hour.
Therefore, if this extreme activity is equally likely on any given day, the LWA surveys, whose observations were composed of 6-hour blocks over 5 days \cite{1538-3881-149-2-65}, then 4-hour blocks over 3 consecutive days \citep{1538-3881-151-2-28}, should have observed pulses of a similar magnitude. 
There is only a $6\%$ chance to randomly select 8 observations which contain no pulses. 

However, the LWA observations were not randomly selected; they took place over consecutive days. This could indicate that the activity level is clustered on timescales of weeks or greater. 
Unfortunately, due to the irregular distribution of our observations we are unable to determine the degree to which GP activity is clustered over timescales of weeks or months. 
Recently, we did a follow-up study utilizing an additional 26 hours of observations, simply monitoring the location of PSR\,B0950+08 with a  $5\sigma$ detection threshold, daily from 2019 September $1^{st}$ to $8^{th}$, as well as the $10^{th}$, $11^{th}$, $14^{th}$, $15^{th}$, and $18^{th}$. We found no GPs at all in these data.
This indicates that there are long periods of quiescence between storms of extreme pulse activity. 
Long periods of quiescence, or low GP detection rates are consistent with the long term study by \cite{2012AJ....144..155S}, who found that 99\% of all pulses were detected during 35 of the 141 days. 

There is however one cautionary comment to make: extremely bright narrow-band signals will often be flagged as RFI, unless they are clearly dispersed at a sizeable dispersion measure. But unhelpfully, the DM of PSR\,B0950$+$08 is very low, and so care should be taken not to discard the brightest GPs via the RFI filtering.

\section{Conclusions}
\label{sec:conclusions}

We report on the observation of a sample of extremely high fluence GPs from PSR\,B0950$+$08. These pulses achieve a fluence brighter than any which have previously been observed at similarly low radio frequencies. Earlier studies show different distributions of GP fluences from this source, either with a much lower rate or with a different distribution shape. Also within our study we find that the rate of GPs is typically constant and unclustered within the few hours of a single observation, but varies by more than an order of magnitude from day to day; it was even seen to go to zero for the better part of a month.

By reprocessing an observation during a very active period, monitoring the position of the pulsar, and subtracting background variability, we were able to extend the sample of detectable GPs to a factor of 5 lower fluence. This revealed that the fluence distribution follows a broken power law, with the break occurring around  a few tens of Jy\,s.
We also found that the shape of the bright end of this distribution was consistent between observations spanning several months, as well as a range in extreme pulse activity. 

We were able to exclude that the bulk of the fluence variations are due to interstellar propagation and magnification effects, and therefore must be attributed to intrinsic behaviour of the pulsar emission. The fine structure in frequency we see also seems more likely to be intrinsic, and is also seen in the GPs of other pulsars, but we cannot strongly exclude that refractive interstellar scintillation plays a role in that.

\section*{Data Availability}

The data underlying this article can be shared on reasonable request to the corresponding author.

\section*{Acknowledgements}

AARTFAAC development and construction was funded by the ERC under the Advanced Investigator grant no. 247295 awarded to Prof. Ralph Wijers, University of Amsterdam; This work was funded by the Netherlands Organisation for Scientific Research under grant no. 184.033.109. We thank The Netherlands Institute  for Radio Astronomy (ASTRON) for support provided in carrying out the commissioning observations. AARTFAAC is maintained and operated jointly by ASTRON and the University of Amsterdam.

We would also like to thank the LOFAR science support for their assistance in obtaining and processing the data used in this work. We use data obtained from LOFAR, the Low Frequency Array designed and constructed by ASTRON, which has facilities in several countries, that are owned by various parties (each with their own funding sources), and that are collectively operated by the International LOFAR Telescope (ILT) foundation under a joint scientific policy.

Additionally, we express gratitude to Joeri van Leeuwen, Anya Bilous, Jason Hessels, Joel Weisberg, and Nina Gusinskaia for their helpful comments on this article, as well as Joanna Rankin and Mathew Bailes for invaluable discussion. And to Jean-Mathias Griessmeier and Louis Bondonneau for sharing illuminating data on the typical pulse behaviour of PSR\,B0950+08.

This research made use of Astropy,\footnote{http://www.astropy.org} a community-developed core Python package for Astronomy\citep{astropy:2013, astropy:2018}, as well as the following:  KERN \citep{molenaar2018kern}, Pandas \citep{mckinney-proc-scipy-2010}, NumPy \citep{2011CSE....13b..22V}, SciPy \citep{citescipy}, and corner.py \citep{corner}. Accordingly, we would like to thank the scientific software development community, without whom this work would not be possible.




\bibliographystyle{mnras}
\bibliography{ExtremePulse} 







\bsp	
\label{lastpage}
\end{document}